\documentclass[aps,prd,twocolumn,showpacs,showkeys,letterpaper]{revtex4}
\usepackage{graphicx}% Include figure files
\usepackage{dcolumn}% Align table columns on decimal point
\usepackage{bm}% bold math

\def\dm31{$\Delta m^2_{31}$}

\def\polfit{\textit{POLfit}}
\def\dmin{$D_{wall}$}
\def\gev{GeV}
\def\mev{MeV}

\begin{document}

%\preprint{APS/123-QED}
%
% Draft version 12
%

\title{Background Study on $\nu_e$ Appearance from a $\nu_\mu$ Beam in Very Long Baseline Neutrino
Oscillation Experiments with a Large Water Cherenkov Detector}

%Lines break automatically or can be forced with \\
\author{C. Yanagisawa\footnote{email: chiaki@nngroup.physics.sunysb.edu; Also at Science Department, BMCC, City University of New York,199 Chambers Street, New York, NY 10007, U.S.A.}, C. K. Jung, P. T. Le\footnote{Currently, Department of Physics and Astronomy,Rutgers, the State University of New Jersey,136 Frelinghuysen Rd, Piscataway, NJ 08854}}
% \altaffiliation[Also at ]{home.}  %  optional
 \affiliation{Department of Physics and Astronomy, Stony Brook University,
Stony Brook, NY 11794-3800, U.S.A.}
% \email{chiaki@nngroup.physics.sunysb.edu}   %optional
\author{B. Viren}
 \affiliation{Department of Physics, Brookhaven National Laboratory,
Upton, NY 11973, U.S.A.\vspace{4mm}}

\date{\today}% It is always \today, today,
             %  but any date may be explicitly specified

\begin{abstract}
There is a growing interest in very long baseline neutrino oscillation
experimentation using accelerator produced neutrino beam as a
machinery to probe the last three unmeasured neutrino oscillation parameters:
the mixing angle $\theta_{13}$, the possible CP violating phase
$\delta_{CP}$ and the mass hierarchy, namely, the sign of $\Delta m^2_{32}$.
Water Cherenkov detectors such as IMB, Kamiokande and Super-Kamiokande
have shown to be very successful at detecting neutrino interactions.
Scaling up this technology may continue to provide the required
performance for the next generation of experiments.
This report presents the latest effort to demonstrate that a next
generation ($>$ 100 kton) water Cherenkov detector can be used
effectively for the rather difficult task of detecting $\nu_e$s from
the neutrino oscillation $\nu_{\mu}\rightarrow\nu_e$ despite the large
expected potential background resulting from $\pi^0$s produced via
neutral current interactions.
\end{abstract}


\maketitle

\section{Introduction}
It has been shown that there are certain advantages in a wideband
neutrino beam over a now traditional narrow-band neutrino beam for
neutrino oscillation experiments, especially for the observation of
$\nu_\mu\rightarrow\nu_{e}$ oscillation, provided that the baseline is
reasonably long (over 1,000 km)~\cite{ref:BNLVLB}.
%
% The major reason for these advantages is that the neutrino oscillation
% parameters assert themselves in different ways and at different
% energies.  

By broadening the range of energies with which multiple
oscillations can be resolved, a richer parameter space can become
accessible than is available to an experiment focused on only the
first oscillation maximum.

In the original paper proposing the use of a wide-band neutrino beam
for a very long baseline neutrino oscillation (VLBNO)
experiment~\cite{ref:BNLVLB} by the Brookhaven National Laboratory
(BNL) neutrino working group (NWG) some simplifying assumptions were
used.  The sensitivities presented in the paper relied on calculations
based on 4-vector level Monte Carlo (MC) simulations, a simple model
for energy resolution and certain assumptions about reconstruction
capabilities.  Namely, a detailed detector simulation for the proposed water
Cherenkov detector was not included.
In addition it was assumed that the signal events were only from
quasi-elastic (QE) charged current (CC) scattering ($\nu_e +
n\rightarrow e^{-} + p$) and the background events were only from
single $\pi^0$ neutral current (NC) interactions $\nu + N\rightarrow
\nu + \pi^{0} + N'$ where $N$ and $N'$ are nucleons.

In order to gain a better insight on this idea of the VLBNO experiment
with a wide-band neutrino beam for $\nu_{\mu}\rightarrow\nu_e$
oscillation, we performed a more sophisticated and elaborate
multivariate likelihood based analysis using a full MC simulation that
included inelastic neutrino interactions, water Cherenkov detector
response, and well-tuned event reconstruction algorithms. This MC
simulation and reconstruction programs were developed and fine-tuned
for the Super-Kamiokande-I experiment (SK-I)~\cite{ref:SKSoft}.

\section{Energy Reconstruction and Event Description}

\begin{figure}
\includegraphics[width=0.45\textwidth]{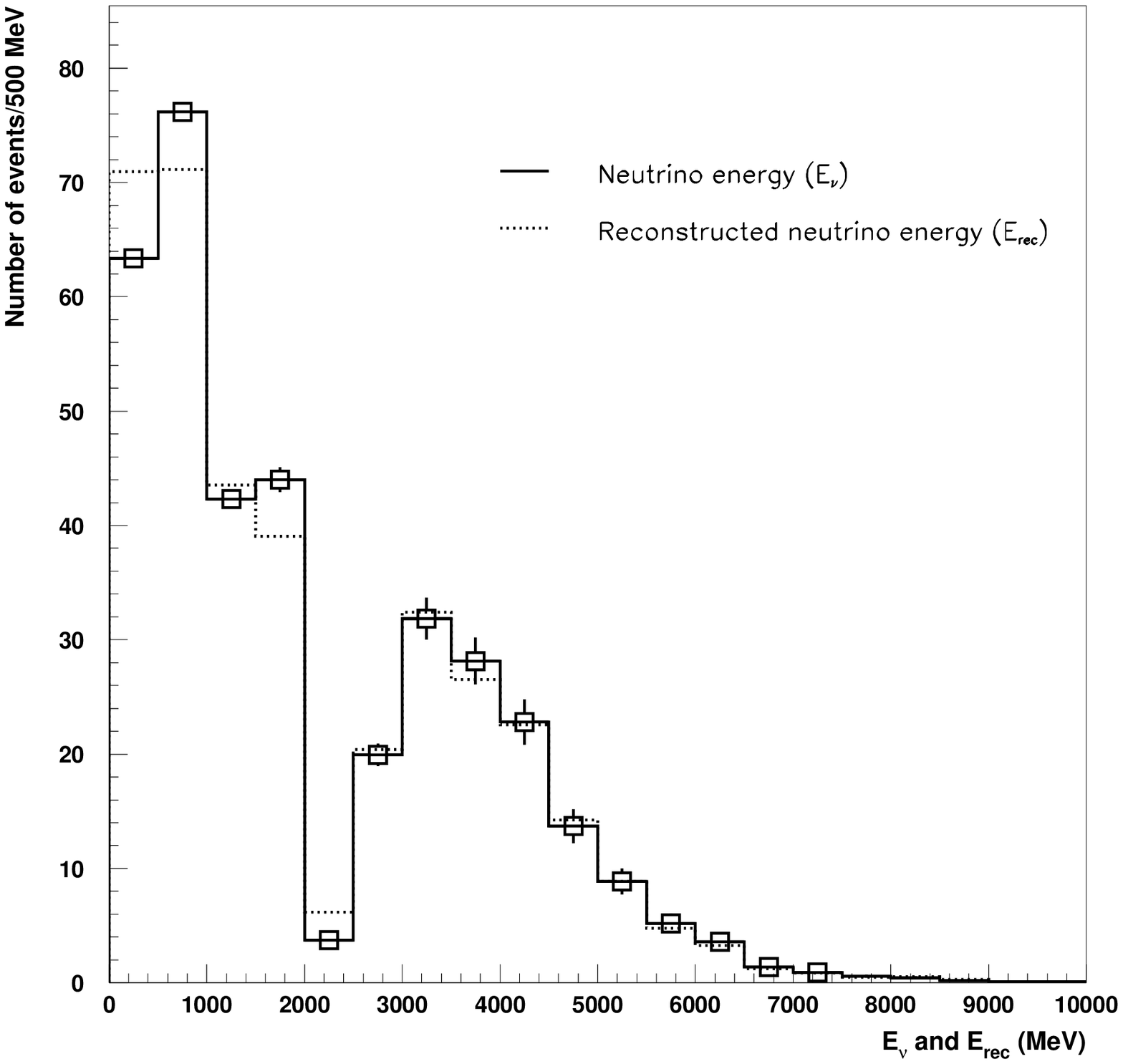}
\includegraphics[width=0.45\textwidth]{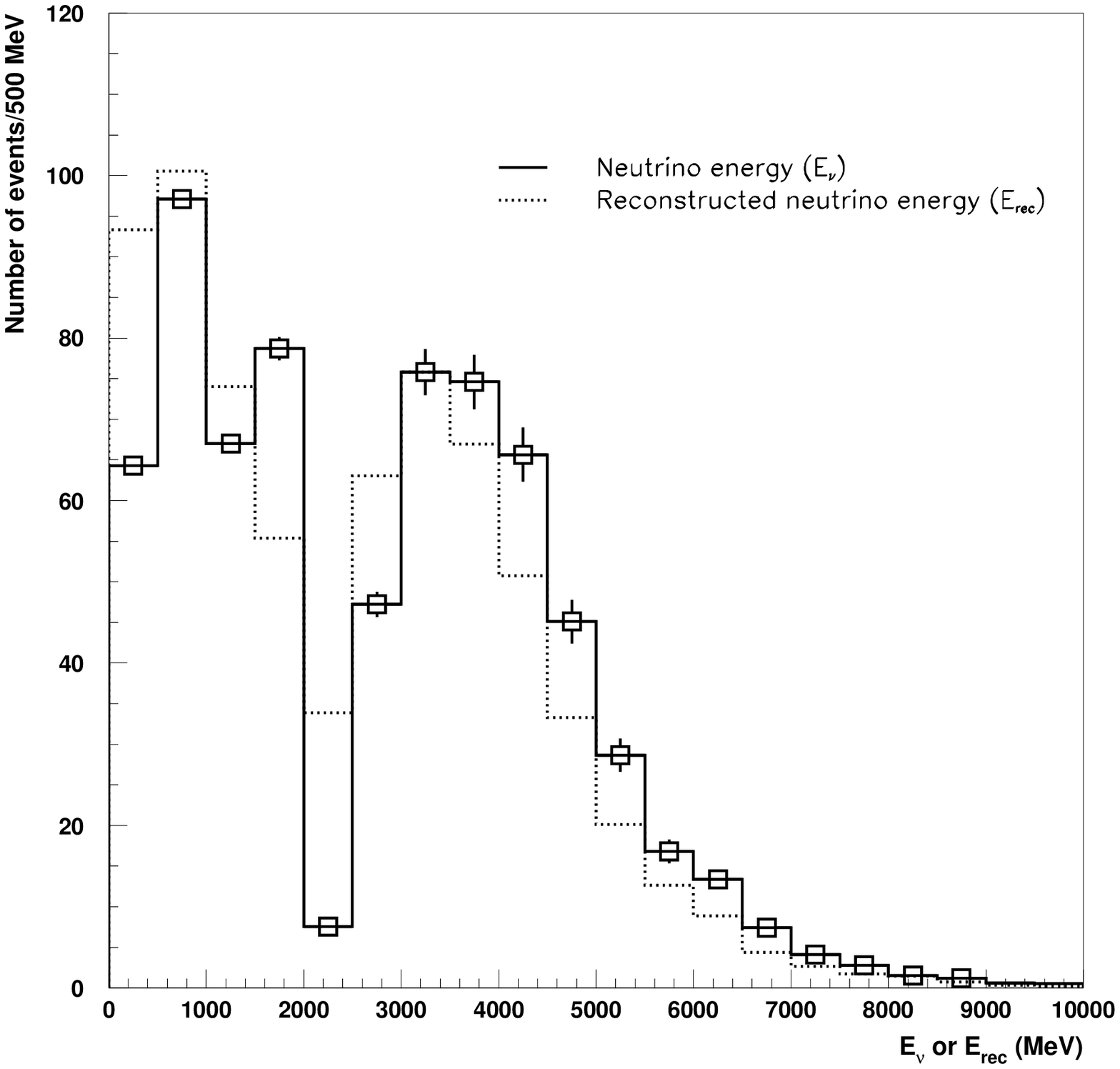}
\caption{\label{fig:erec} The distributions of neutrino energy (solid
  lines) and reconstructed neutrino energy (dotted lines) of single
  ring e-like events originating from CC QE interactions (top figure)
  and from all CC interactions (bottom figure). Note that a dip at
  about 2 \gev{} is due to $\nu_{\mu}\rightarrow\nu_e$ neutrino
  oscillation.  The open square boxes with error bars indicate
  statistical uncertainties.}
\end{figure}

For the discussions in this paper, we calculate a reconstructed
neutrino energy using the formula

%\[E_{rec}=m_{N}E_{e}/[{m_{N}-(1-cos\theta_{e})E_{e}}]\]
\[E_{rec}=\frac{m_{N}E_{e}}{m_{N}-(1-cos\theta_{e})E_{e}},\]

\noindent where $m_{N}$, $E_e$ and $\theta_e$ are the nucleon mass,
the recoil electron energy and the scattering angle of the recoil
electron with respect to the incident neutrino beam, respectively. In
a strict sense, this quantity represents the incident neutrino energy
only when the event is produced by CC QE scattering and the Fermi
motion of target nucleons is ignored. Nonetheless, we show in
Figure~\ref{fig:erec} the neutrino energy and $E_{rec}$ for
CC QE (top) and all CC (bottom) events to compare the two energy 
distributions for the two classes of CC events.  Although $E_{rec}$ for the CC
events does not reproduce the incident neutrino spectrum as well as it
does for CC QE events, it still reproduces it quite well.

The initial event selection is made to maximize the number of
interactions that are consistent with coming from $\nu_e$ appearing 
% added 7/22/10
at the far end of the baseline
in the beam while minimizing events
such as $\nu_\mu$ CC or any events from NC interactions.
Needless to say, the clearest event signature comes from $\nu_e$ CC QE 
interactions. Inelastic $\nu_e$ CC events also carry information about 
oscillations but are more difficult to cleanly select and have somewhat 
worse energy resolution as shown in Figure~\ref{fig:erec}.  
Regardless of the interaction, a recoil proton will rarely be
above its Cherenkov threshold of about 1.25 \gev{}$/c$ and will not
contribute notably to the event signature.
For these reasons, a $\nu_e$ appearance candidate event will be
initially selected if its signature is consistent with being a
Cherenkov radiation pattern from a single electron (termed \textit{single
  ring, e-like}).

A $\nu_\mu$ CC QE event will usually be distinguishable from one
arising from a $\nu_e$ CC QE interaction as the resulting Cherenkov
ring produced by $\nu_\mu$ will have a sharper rise and fall in the amount of
Cherenkov light at the edges of the ring than one due to an electron.
This is because a muon presents a relatively straight line source of
Cherenkov light while an electron initiates an electromagnetic shower.
This shower consists of a distribution of track directions due to the
number of particles in the shower and that they experience larger
multiple scattering.  As the muon energy approaches the Cherenkov
threshold there is an increasing chance that multiple scattering and
the collapsing Cherenkov light cone will smooth the rise of the edge
of the ring and potentially mimic an electron.

Likewise, inelastic and NC interactions with a non-zero pion
multiplicity can have additional light that is not consistent with the
pattern of a single e-like ring.  Depending on the exact event
topology these events are either easily ruled out or present
challenging backgrounds.  Charged pions above their Cherenkov
threshold of 160 \mev{}/c produce a $\mu$-like ring and are rejected.
Events with visible light from both pions and the primary charged lepton are
strongly rejected based on ring counting.

Events with a final state consisting of only a single neutral pion
present a particular challenge.  The $\pi^0$ decays to a pair of gamma
rays, each of which initiate an electromagnetic shower and produce an
e-like ring of varying intensity and direction.  These $\pi^0$ events
can become background depending on how the $\pi^0$ decays to two gamma
rays.  At one extreme, the decay is symmetric such that the gamma rays
have similar energies.  If the original $\pi^0$ is boosted enough, the two
gamma rays are nearly collinear and produce mostly overlapping rings.
They can then be impossible to be distinguished single electron events
with an energy equal to the sum of the two gammas.  At the other
extreme, a highly asymmetric $\pi^0$ decay results in one strongly
boosted and one strongly retarded gamma ray.  In the lab frame, the
higher energy gamma will produce a single e-like ring while the other
may produce no discernible light.

The events with a $\pi^0$ in the final state can come from two
sources: (1) NC interactions and (2) charge exchanges of charged pions
inside the target nucleus or with an oxygen or a hydrogen nucleus while 
traveling in water.

Finally, there are events from electron neutrino interactions that are
%not from $\nu_e$ that appeared in the beam.  Rather, there is an
% modified 7/22/10
not from $\nu_e$ that appeared in the beam from neutrino oscillation 
$\nu_\mu\rightarrow\nu_e$.  Rather, there is an
intrinsic $\nu_e$ component in the unoscillated neutrino beam
originating from muon or kaon decays.  The beam used in this study has
an intrinsic $\nu_e$ component which is 0.7\% of the $\nu_\mu$ flux.
The background events from these neutrinos are irreducible.

In this analysis, to classify the events for the initial selection,
the reconstruction algorithms used for the SK-I atmospheric neutrino
analysis are employed.  In addition, a special algorithm called
\polfit{}~\cite{ref:Barszczek} is used.  This algorithm has been found
to be extremely useful for removing single $\pi^0$ background events
in samples containing events that are classified as single-ring and
e-like~\cite{ref:polfit} by the standard SK-1 reconstruction. This
paper describes how information provided by \polfit{}, together with
other useful variables, can be effectively used to significantly
reduce the $\pi^0$ background while retaining a sufficient efficiency
for the signal.

\section{Monte Carlo event sample}
The Monte Carlo event sample used in this study was 
originally produced by the SK-I collaboration to
simulate atmospheric neutrino events detected by the SK-I detector
and
corresponds to about 100 years of the exposure. The detailed description
of the Monte Carlo generation as well as the event reconstruction 
is described elsewhere~\cite{ref:ATMFull}. 

The energy spectra
for $\nu_\mu$ and $\nu_e$ produced in the Earth's atmosphere are 
different from those in the
neutrino beam proposed by the BNL NWG.
To account for this, each $\nu_\mu$ event is given a weight based on
the neutrino energy such that the shape of the resulting spectrum
matches that expected from the beam.  An additional normalization is
applied so that the total number of $\nu_\mu$ CC QE events is as
expected.  

Assuming no oscillations, it is expected that there will be 12,000
such events for a detector with SK-I efficiencies and fiducial volume
of 500 kton (22.2 $\times$ SK-I) which is placed 2,540 km from the
neutrino production target and for five years of running as described
in the paper by BNL NWG~\cite{ref:BNLVLB}.  When oscillations are
applied, this sample is weighted by the oscillation probability taking
into account full three-neutrino mixing and matter effects.

The oscillated $\nu_e$ sample is prepared by similarly weighting and
normalizing the $\nu_e$ atmospheric MC events to what is expected from
the $\nu_\mu$ component of the beam, that is, under the assumption of
100\% $\nu_\mu \to \nu_e$ oscillation.  Actual oscillation
parameters are then applied to this sample by weighting to the
appropriate oscillation probability.  The intrinsic $\nu_e$ component
in the beam is taken to be the shape predicted by the beam MC simulation
and is normalized to be 0.7\% that of the $\nu_\mu$ component.
% addition for PRDv12 about nue oscillation
To include the effect of neutrino oscillation, the contribution of the
$\nu_e$ component is calculated only when $\nu_e$ survives as $\nu_e$ at
the water Cherenkov detector placed at a given baseline.

% PRDv11
%The anti-$\nu$ components of the beam are ignored in this study.  Both
%($\bar\nu_\mu$ and $\bar\nu_e$) are small in number compared to their
%$\nu$ partners in the beam.
% PRDv12
The expected numbers of the events to be observed from the intrinsic
anti-$\nu_\mu$ component of the beam are estimated to be 2.2\% and 2.5\%
of the expected numbers of the background events from the $\nu_\mu$ 
(background-1) for the baselines of 2,540 km and of 1,480 km, respectively
(see the next section for the description about the baseline). The expected 
numbers of the events to be observed from anti-$\nu_e$ component of the beam
are estimated to be less than 4.5\% and 4.3\% of the expected numbers of the
background events from the intrinsic $\nu_e$ component of the beam
(background-2) for the baselines of 2,540 km and of 1,480 km, respectively.
Since these contributions are much smaller than the events from 
the background-1 and the background-2, although not negligible, we will not
take into account them further in this report.

To select MC $\nu_e$ appearance candidate events an initial set of
cuts based on the standard SK-I codes is used:
(a) there is one and only one e-like ring, 
(b) the reconstructed event vertex is at least 2 m from any inner
photomultiplier (PMT) surface and 
(c) there is no activity in the outer (veto) detector.

In addition to these standard cuts, events produced by neutrinos with
energies greater than 10 \gev{} are ignored due to lack of statistics in
the atmospheric MC sample.  The beam has a tail that extends up to 15
\gev{} but its contribution to the background is negligible as will be
shown below.

\section{Neutrino Oscillation Probabilities}

The oscillation probabilities applied to the components of the beam
neutrino flux use the following parameters:

\begin{center}
\begin{tabular}{rcl}
$\Delta m^{2}_{21}$   & = & $7.3\times 10^{-5} eV^2$, \\
$\Delta m^{2}_{31}$   & = & $2.5\times 10^{-3} eV^2$, \\
$\sin^{2}2\theta_{12}$ & = & $0.86$, \\
$\sin^{2}2\theta_{23}$ & = & $1.0$, \\
$\sin^{2}2\theta_{13}$ & = & $0.04$, and  \\
$\delta_{CP}$        & = & $0^{\circ},\pm 45^{\circ},\pm 135^{\circ}$.
\end{tabular}
\end{center}

% %
% $\Delta m^{2}_{21} = 7.3\times 10^{-5} eV^2$,
% %
% $\Delta m^{2}_{31} = 2.5\times 10^{-3} eV^2$,
% %
% sin$^{2}\theta_{12} = 0.86$, 
% %
% sin$^{2}\theta_{23} = 1.0$,
% %
% sin$^{2}\theta_{13} = 0.04$, and 
% %
% $\delta_{CP} = 0^{\circ},\pm 45^{\circ},\pm 135^{\circ}$.
% %

Note that the sign convention of $\delta_{CP}$ follows that of the
paper by BNL NWG~\cite{ref:BNLVLB}. Unless otherwise stated, in this
paper the value of $\delta_{CP}$ is +45$^\circ$ and the baseline is
2,540 km which corresponds to the distance from BNL to Homestake Mine
in South Dakota. Later we will describe the results with different
values for $\delta_{CP}$ as well as with the baseline of 1,480 km
which corresponds to the distance from Fermilab to Henderson Mine in
Colorado.  
The result with the baseline of 1,480 km was for
the UNO
detector proposed first in 1999 and later to be built in Henderson
Mine. See the references~\cite{ref:uno} for the detail of the UNO detector.
 These results are similar to what is expected for the
Fermilab to Homestake baseline of 1,300km.
This shorter baseline is considered by applying inverse-square scaling
of the flux and recalculation of oscillation probabilities.

The neutrino oscillation probabilities were calculated by the
\textit{nuosc}~\cite{ref:nuosc} package.  The calculator assumes
3-$\nu$ oscillation with possible CP symmetry violation and matter
effects.  The matter densities can be vacuum, non-zero constant or
stepwise-constant matter densities.  The non-constant
PREM~\cite{ref:prem} and arbitrary user provided density profiles are
also supported.  It calculates the probabilities either analytically
for constant matter densities or for any density profile it can use a
5th order adaptive Runge-Kutta stepper.  The package has been
validated against published calculations~\cite{ref:irina}.

\section{POLFIT}
In this section we briefly describe the $\pi^0$ fitter called \textit{Pattern Of
Light fit} (\polfit{})~\cite{ref:Barszczek}. 
\polfit{} is applied to events which are initially classified as
single-ring and e-like by the standard SK-I analysis codes in order to
identify $\pi^0$ background events in this event sample.
It assumes the event is a $\pi^0$ event and that the ring found by
the standard codes is one of the gammas from the $\pi^0$ decay.  It
will then determine the direction and energy of a secondary gamma
that, along with the primary one, is most consistent with the pattern
of light collected by the PMTs.  \polfit{} produces two possible
secondary gammas by employing two algorithms.  One is optimized
assuming the secondary ring is overlapping with the primary ring
(forward-algorithm).  
%The other is optimized assuming the two rings
%are non-overlapping (wide-algorithm).
The other is optimized to find a second ring in the wider angular
region (wide-algorithm).

In each algorithm, the pattern of light expected from the full $\pi^0$
decay is calculated from a set of templates and compared to the
collected light and a likelihood is formed.  These templates are
pre-determined using the detector simulation.  The energy and
direction phase-space of the secondary gamma is searched to find
the point that leads to the maximum likelihood.
Each algorithm supplies the direction of two photons, their energies,
the two photon invariant mass, and the maximum likelihood value obtained.
It was found that the wide-algorithm method more reliably finds a real
second ring more often than the forward-algorithm.  Therefore, in this
report we use mostly the information provided by the wide-algorithm
unless otherwise stated.

\begin{figure}
\includegraphics[width=0.45\textwidth]{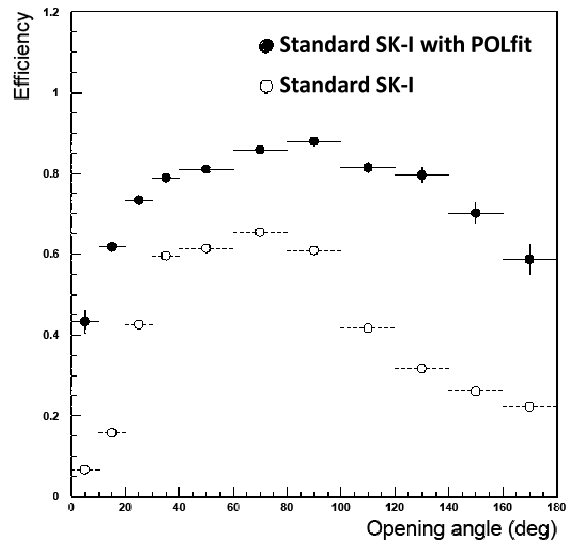}
\caption{\label{fig:pi0effic} $\pi^0$ reconstruction efficiency with the standard SK-I codes (open circles) and after accepting additional events with missing gamma found by  \polfit{} (solid circles).}
\end{figure}

To demonstrate the power of \polfit{}, Figure~\ref{fig:pi0effic} shows
the relative $\pi^0$ reconstruction efficiencies with (solid circles)
and without (open circles) \polfit{} employed as a function of the
opening angle between the two reconstructed gammas from the $\pi^0$
decay in the lab frame and for single $\pi^0$ events produced by the
NC interactions.
To be accepted in this sample, the event must have a two-photon invariant mass
($m_{\gamma\gamma}$) within $2\sigma$ of the expected value given
the event came from a $\pi^0$.
Here, the neutrino energy spectrum is that of the original SK-I
atmospheric muon neutrinos.
In the case without \polfit{}, only events identified as having two
e-like rings by the standard SK-I software can be candidates to be a
$\pi^0$.  With \polfit{} information, $\pi^0$ events that are
classified as single-ring by the standard selection may now be
selected.

Depletion of $\pi^0$ detection efficiencies at a small opening angle
is due to overlap of the two e-like rings.  At large opening angle it
is due to the second ring being too faint.   As clearly seen in
Figure~\ref{fig:pi0effic}, \polfit{} improves the $\pi^0$ reconstruction
efficiency significantly.

\section{Useful variables to distinguish the signal from the background}

A cut on $m_{\gamma\gamma}$ reduces background from single $\pi^0$ NC
events but it also reduces signal efficiency somewhat.  Alone, it is
not enough to reduce the background to a level comparable to the
oscillation signal.
It is, therefore, desirable to find additional distinguishing features.

In this section, nine such features (variables) including
$m_{\gamma\gamma}$ are described.  Distributions of these variables
are shown for events that are oscillated $\nu_e$ CC (labeled
``signal'' and with solid lines) and NC, $\nu_\mu$ CC
and beam-$\nu_e$ CC events (``background'', dotted lines).  The
distributions are plotted for different $E_{rec}$ regions in steps of
0.5 \gev{} from 0 to 2 \gev{}, 2 \gev{} $\le E_{rec} <$ 3 \gev{} and
$E_{rec} \ge$ 3 \gev{}.

In general, at lower energies (sub-\gev{} range), the background is
largely from the misidentified single-$\pi^0$ events discussed above.
At higher energies, non-QE interactions become dominant leading to
additional ways for background to mimic $\nu_e$ CC QE events.

\subsection{Reconstructed $\pi^0$ mass ($m_{\gamma\gamma}$)}
\begin{figure}
\includegraphics[width=0.45\textwidth]{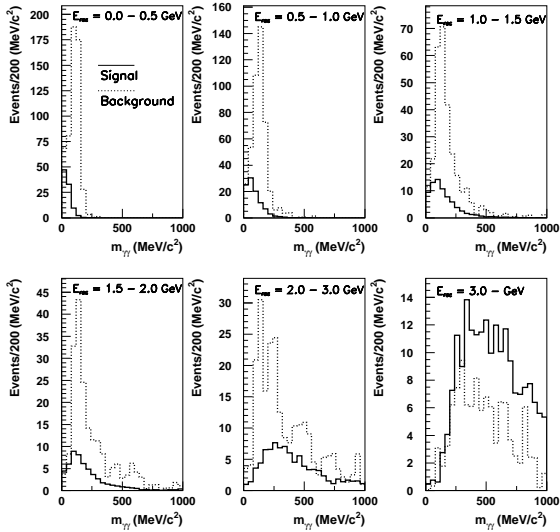}
\caption{\label{fig:pi0mass} The distributions of reconstructed $\pi^0$ mass
of single e-like ring events for signal (solid line) and background
(dotted line).}
\end{figure}

As shown above, this variable is useful to remove background events
that come from single $\pi^0$ NC production.
%%%%%        In Figure~\ref{fig:pi0mass_efrac} the distributions of the
In Figure~\ref{fig:pi0mass} the distributions of $m_{\gamma\gamma}$
are shown.
In the lower reconstructed neutrino energy region ($E_{rec}\leq$ 2
\gev{}), the distributions show a prominent $\pi^0$ peak.  At higher
energies, the contribution from multi-pion production increases and
\polfit{} begins to reconstruct the correct second gamma poorly or
returns an arbitrary result that is not associated with any physical
particle.  Because of this, the $\pi^0$ peak disappears in the higher
energy region.

%%%% This will be split
%\begin{figure}
%\includegraphics[width=0.45\textwidth]{pi0massprdBW}
%\includegraphics[width=0.45\textwidth]{efracprdBW}
%\caption{\label{fig:pi0mass_efrac} Left figures: The distributions of
%reconstructed
%$\pi^0$ mass of single e-like ring events for the signal (solid line) and
%the background (dotted line). Right figures: The distributions of the energy
%fraction of the second ring found by \polfit{} in 1-ring events for the signal
%(solid line) and the background (dotted line).}
%\end{figure}
%%%%

\subsection{Fraction of energy in the second ring (E$_{frac}$)}
\begin{figure}
\includegraphics[width=0.45\textwidth]{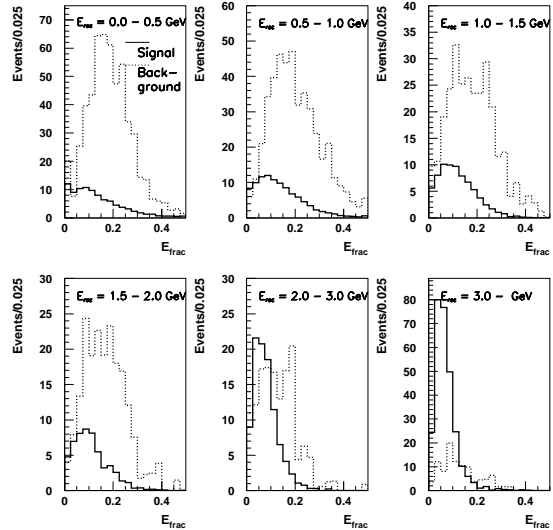}
\caption{\label{fig:efrac} The distributions of the energy fraction of the
second ring found by \polfit{} in 1-ring events for the signal (solid line) and
background (dotted line).}
\end{figure}

In general, the second e-like ring found by \polfit{} has less energy
than the one found by the standard SK-I
reconstruction software.  
Furthermore, \polfit{} must find a second ring and if the event does
not in fact have one, this wrongly reconstructed ring 
tends to have much less energy than the primary one.
%%% well seen in Figure~\ref{fig:pi0mass_efrac} (right figures) where the
This effect is shown in Figure~\ref{fig:efrac} where the ratios of
the energy of the second ring to the sum of two ring energies are
plotted.  

\subsection{Difference in log $\pi^0$ likelihood ratio ($\Delta$log $\pi^0$-lh)}

\begin{figure}
\includegraphics[width=0.45\textwidth]{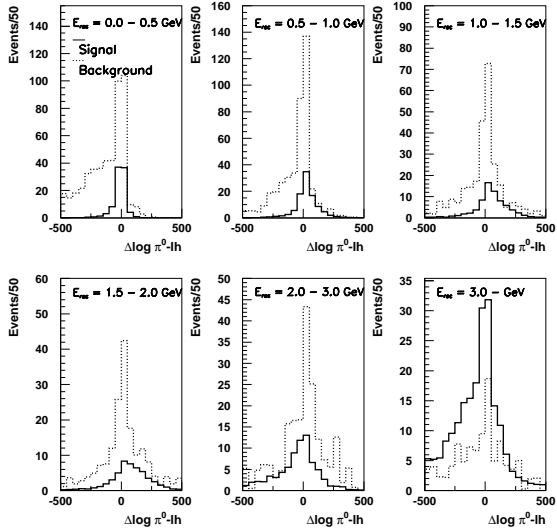}
\caption{\label{fig:dpi0likelihood} The distributions of the difference in 
log $\pi^0$ likelihood between the two algorithms. The larger this variable,
the more likely an event is the $\pi^0$ background. The distributions in solid
lines are for the signal and those in dotted lines are for the background.}
\end{figure}

As mentioned above, \polfit{} employs two algorithms, the wide- and
forward-algorithm. Each provides a log-likelihood that the event is a
$\pi^0$.
Signal events actually contain only a single ring.  Nevertheless, with
such events, \polfit{} will return a second, fake ring.  In such cases,
both algorithms tend to place this fake ring in the vicinity of the
single primary ring.
Therefore, the two algorithms tend to find similar fake rings.
This makes the two likelihoods similar to each other and the
log-likelihood ratio (difference between two log-likelihoods) tends to
be symmetric around zero.
On the other hand, in the case of a $\pi^0$ background event the
secondary photon from a $\pi^0$ decay is not necessarily confined in
the direction of the primary photon.  This makes the log-likelihood
ratio asymmetric with a long left tail, especially at lower
energies. The difference is
significant in the lower energy region as shown in
% Figure~\ref{fig:dpi0likelihood_costh} (left figure). However, this
Figure~\ref{fig:dpi0likelihood}. However, this trend
changes above $E_{rec}$ = 1.0 \gev{} where the contribution from
multi-pion production starts to increase and dilutes this effect.

%%%%% To be split
%\begin{figure}
%\includegraphics[width=0.45\textwidth]{dlikelihoodprdBW}
%\includegraphics[width=0.45\textwidth]{costhprdBW}
%\caption{\label{fig:dpi0likelihood_costh} Left figures: The distributions of
%the difference in log $\pi^0$ likelihood between the two \polfit{} algorithms. The
%larger this quantity, the more likely an event is a $\pi^0$ background. The
%distributions in solid line are for the signal and those in dotted line are for
%the background. Right figures: The distributions of the directional cosine of 
%the primary e-like ring with respect to the neutrino beam direction. The
%distributions in solid are for the signal and those in dotted are for the
%background.}
%\end{figure}

\subsection{Direction cosine of the e-like ring (cos$\theta$)}

\begin{figure}
\includegraphics[width=0.45\textwidth]{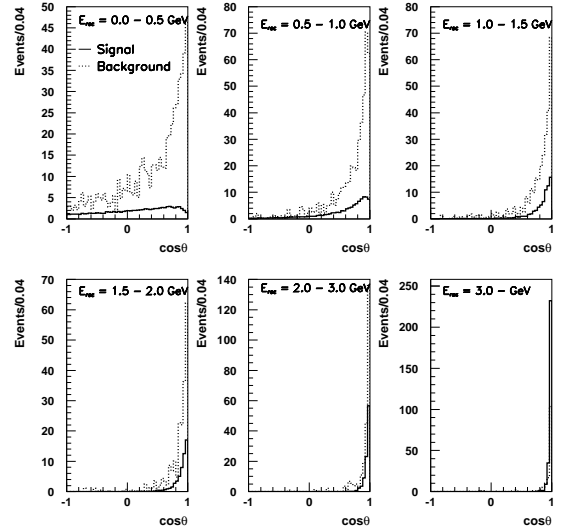}
\caption{\label{fig:costh} The distributions of the directional cosine of the
primary e-like ring with respect to the neutrino beam direction. The
distributions in solid lines are for the signal and that in dotted lines are 
for the background.}
\end{figure}

The direction cosine of the primary e-like ring with respect to the
neutrino beam is a good discriminator to separate the signal from the
background.  Its distributions are shown in Figure~\ref{fig:costh}.
This variable does not depend on the information from \polfit{}. 

As the energy of the neutrino decreases, the outgoing electron
direction is distributed over a wider angle with respect to the
neutrino direction.  As the energy becomes comparable to that of Fermi
motion, the distribution is further widened because the momentum of the
target nucleon becomes significant and will tend to randomize the
electrons direction.  This latter effect is less for the more massive
$\pi^0$.

% There is also a systematic sampling effect, caused by applying the
% $E_{rec}$ formula to $\pi^0$ events, which contributes to the
% distributions peaking towards unity.  It occurs because the assumption
% that the outgoing visible particle is an electron is violated for these
% events.  As the angle between the decay-gamma ring and the neutrino
% direction increases the probability increases that the $E_{rec}$
% formula will produce an unphysical energy and the event will be
% removed from the energy bin.

% These effects lead to the result that the $\pi^0$ background events,
% as binned in reconstructed energy, appear to point in the neutrino
% direction better than do the $\nu_e$ CC QE events at low energy.

% Due to the Fermi motion of the target nucleon,
% the recoil electron in a signal event is scattered more than
% the $\pi^0$ in a background event in the lower neutrino energy region.
% Since we require that a candidate signal event be of a single e-like
% event, the second lower energy photon in a candidate event from the
% $\pi^0$ background is missed and the majority of the $\pi^0$ energy is
% carried by the primary e-like ring that tends to go in the forward
% direction. This is another reason why the
% electron in a signal event is scattered more than the primary photon in
% a background event, as clearly seen in
%	Figure~\ref{fig:dlikelihood_costh} (right figures).

\subsection{Total charge to ring energy ratio (Q/E)}

\begin{figure}
\includegraphics[width=0.45\textwidth]{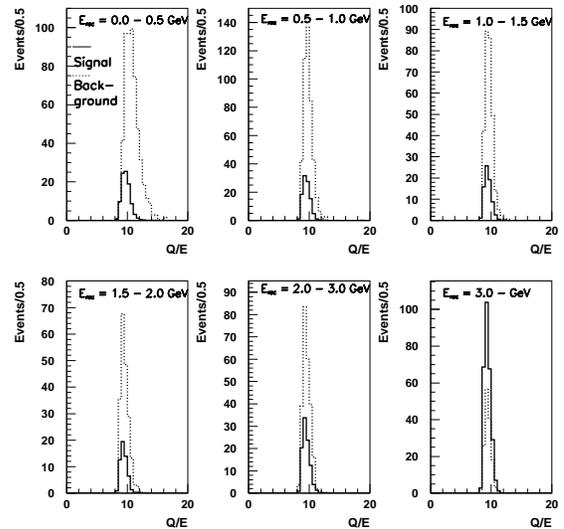}
\caption{\label{fig:poa} The distributions of the ratio, the total charge in 
PE to the ring energy in \mev{} are shown. The distributions in solid line are 
for the signal and those in dotted line are for the background.}
\end{figure}

Light collected in an event that is not consistent with the single
ring found by the standard SK-I codes is an indication of
unreconstructed or missed particles and thus a non-QE interaction.  A
measure of this light is the ratio of the total charge, in unit of PE 
(photoelectron), collected by all PMTs to the energy associated with 
the primary reconstructed ring in \mev{}.  Some background events where 
only one ring is found are expected to produce some light which is not
identified as a ring.  The distributions of this variable are shown in
Figure~\ref{fig:poa}.  The separation is most pronounced at lower
reconstructed energy where any $\pi^0$s are likely to be less boosted
and not put all their light in the forward direction.  Thus in an
asymmetric $\pi^0$ decay the second ring tends to be weaker than the
standard SK codes can detect.

%%% To be split
%\begin{figure}
%\includegraphics[width=0.45\textwidth]{ptoteprdBW}
%\includegraphics[width=0.45\textwidth]{pid2prdBW}
%\caption{\label{fig:poa_elike} Left figures: The distributions of the ratio, 
%the total charge in pe to the ring energy in \mev{} are shown. The distributions
%in solid are for the signal and those in dotted line are for the background.
%Right figure: The distributions of the log-likelihood ratio
%between e-like and $\mu$-like of the primary e-like ring. The distributions in
%solid line are for the signal and those in dotted line are for the background.}
%\end{figure}
%%%

\subsection{log particle-identity-likelihood (log pid-lh)}
\begin{figure}
\includegraphics[width=0.45\textwidth]{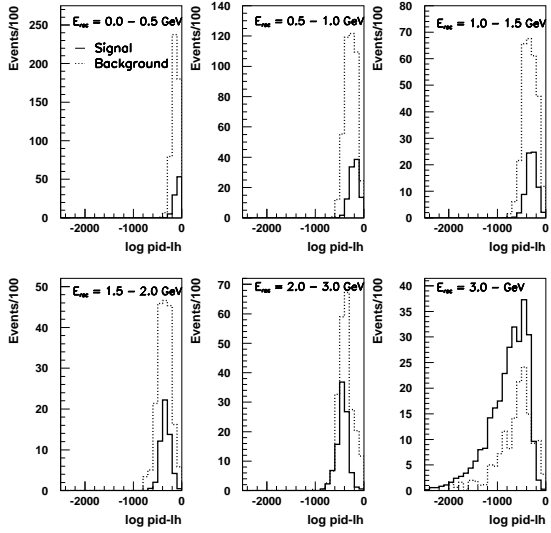}
\caption{\label{fig:elike} The distributions of the log-likelihood ratio
between e-like and $\mu$-like of the primary e-like ring. The distributions in
solid line are for the signal and those in dotted line are for the background.}
\end{figure}

The standard SK-I reconstruction software provides likelihoods of
a Cherenkov ring to be e-like ($L_{e}$) and $\mu$-like
($L_{\mu}$). From these two likelihoods, the logarithm of the ratio
of likelihoods, $\log(L_{\mu}/L_{e})$ (log pid-lh) is used as a
good measure to separate electrons from muons. The more negative
log pid-lh is, the more e-like a Cherenkov ring is.  
When two e-like rings, such as the photons from a $\pi^0$ decay,
overlap and reconstruct as a single ring they can produce a log pid-lh
that is more e-like than a single electron at a comparable energy
would have.  Figure~\ref{fig:elike} shows this small but noticeable
effect.

\subsection{log $\pi^0$-likelihood (log $\pi^0$-lh)}

\begin{figure}
\includegraphics[width=0.45\textwidth]{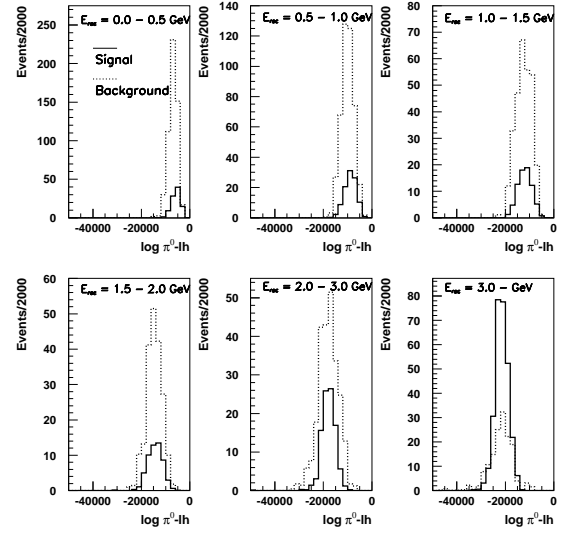}
\caption{\label{fig:pi0like} The distributions of the log $\pi^0$-likelihood
of single e-like ring events. The distribution in
solid line is for the signal and that in dotted line is for the background.}
\end{figure}

\polfit{} provides the logarithm of a likelihood that the event
consists of a single $\pi^0$.  For events that actually do consist of
a single $\pi^0$ this variable is expected to be smaller (more
negative), namely more $\pi^0$-like, than that for a signal
event. This trend shows up in lower energy region ($E_{rec}\leq$1
\gev{}) as shown in
% Figure~\ref{fig:pi0like_Cangle} (left figures). At higher energy 
Figure~\ref{fig:pi0like}. In the higher energy 
region, the distribution tends to be narrower for the signal events.

%%% To be split
%\begin{figure}
%\includegraphics[width=0.45\textwidth]{pi0likelihoodprdBW}
%\includegraphics[width=0.45\textwidth]{angeprdBW}
%\caption{\label{fig:pi0like_Cangle} Left figures: The distributions of the
%$\pi^0$-likelihood of the primary e-like ring. The distributions in
%solid line are for the signal and those in dotted line are for the background.
%Right figures: The distributions of the measured Cherenkov angle of
%the primary e-like ring. The distributions in solid line are for the signal and
%those in dotted line are for the background.}
%\end{figure}

\subsection{Cherenkov angle (Cangle)}
\begin{figure}
\includegraphics[width=0.45\textwidth]{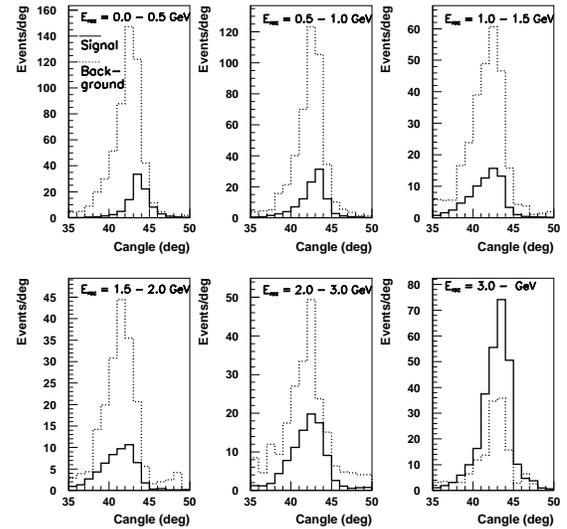}
\caption{\label{fig:Cangle}The distributions of the measured Cherenkov angle of
the primary e-like ring. The distribution in solid line is for the signal and
that in dotted line is for the background.}
\end{figure}

The distribution of reconstructed Cherenkov angle is expected to be
different between the signal and background events and it depends on
whether there is overlap between two Cherenkov rings and on the energy
% of the primary ring. In Figure~\ref{fig:pi0like_Cangle} (right figures)
of the primary ring. In Figure~\ref{fig:Cangle}
the Cherenkov angle distributions for different
$E_{rec}$ regions are shown. The shape of
the distribution for the signal events differs from that for the
background events in most of the energy regions, although degrees of
differences vary from energy region to region.

\subsection{Log ring-count likelihood ratio ($\Delta$log ring-lh)}
\begin{figure}
\includegraphics[width=0.45\textwidth]{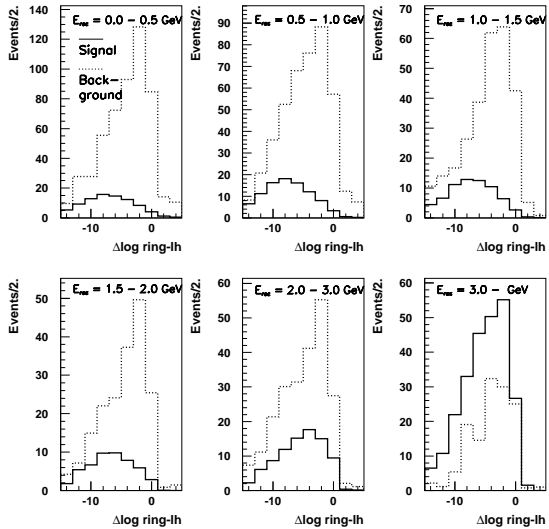}
\caption{\label{fig:dlfct}The distributions of log-likelihood ratio of
  two hypotheses on the number of Cherenkov ring in an event: there is
  only one Cherenkov ring or there are more. If the log-likelihood
  ratio is negative, the event is less likely to have an additional
  ring. The distributions in solid line are for the signal and that in
  dotted line are for the background.}
\end{figure}

To decide how many Cherenkov rings there are in an event, two
hypotheses are compared: the first hypothesis is that there is only
one Cherenkov ring and the second is that there is an additional ring
in the event.  The comparison is made using
log-likelihood ratio, calculated by the standard SK-I codes, of the
two hypotheses. In Figure~\ref{fig:dlfct} this log-likelihood ratio is
plotted for different $E_{rec}$ regions. The shape of the distribution
for the signal events differs from that for the background events in
most of the energy regions, although degrees of differences again vary
over the different energy regions.

\section{Discriminator: Likelihood function and likelihood ratio}
A series of cuts on the variables described in the previous section
can reduce the background well. However, too many cuts in series
reduce the signal selection efficiency. The problem for this analysis
is that none of the variables alone can separate the signal from the
background powerfully. Thus, this situation is well suited for a
multivariate analysis where a likelihood function is defined and
utilizes all variables that have noticeable distinguishing power. We
have shown that the nine variables described above are distributed
differently depending on the source of the events (signal or
background), although the differences may not be large. As we will
see, an accumulation of relatively small differences, when the
correlations among the variables are small, can make a bigger
difference.

From each distribution shown in
Figures~\ref{fig:pi0mass}-\ref{fig:dlfct}, a probability is calculated
for an event to have a value of the corresponding variable $i$ as
signal $p^{s}_{i}$ and as background $p^{b}_{i}$. Then the
log-likelihood is defined by $\log(L_{s}) =
\Sigma_{i=1,..,9}\log(p^{s}_{i})$ as signal and by
$\log(L_{b})=\Sigma_{i=1,..,9}\log(p^{b}_{i})$ as background.  These
probabilities are calculated in the same bins of $E_{rec}$ as used in
Figures~\ref{fig:pi0mass}-\ref{fig:dlfct}.

\begin{figure}
\includegraphics[width=0.45\textwidth]{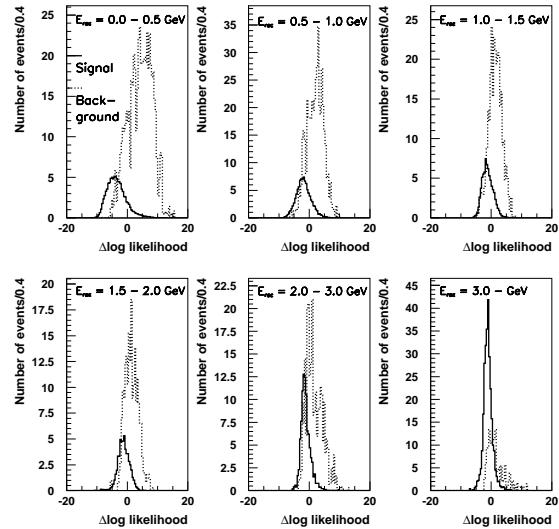}
\caption{\label{fig:dlogL}The distributions of log-likelihood ratio of
two hypotheses on the origin of events: signal vs. background.
The distributions in solid line are for the signal and those in dotted line
are for the background.}
\end{figure}

Then the difference between two log-likelihoods, $\Delta \log(L) =
\log(L_{b})-\log(L_{s})$, is calculated to decide  whether to
accept or reject the event. Figure~\ref{fig:dlogL} shows $\Delta
\log(L)$ distributions for the different $E_{rec}$ regions for the
signal events (solid line) and the background events (dotted
line). The smaller $\Delta \log(L)$ is, the more likely an event is a
signal event. It is clearly seen that the $\Delta \log(L)$
distribution of the signal events differ significantly from that of
background events over wide range of $E_{rec}$.

\section{Reconstructed neutrino energy distributions with a cut on $\Delta \log(L)$}

The final acceptance of an event is determined by its value of $\Delta \log(L)$.
As the distributions of the nine variables used to define the
likelihood depend on $E_{rec}$, so does the distribution of $\Delta
\log(L)$.  Therefore, in order not to change the energy spectrum
unnecessarily, we adopt a strategy to keep the signal detection
efficiency constant over a wide range of $E_{rec}$ by changing the
$\Delta \log(L)$ cut according to $E_{rec}$.  The cut is selected to
retain 40\% of the signal that passes the standard SK-I cuts for all
$E_{rec}$.

\begin{figure}
%PRDv11
%\includegraphics[width=0.45\textwidth]{ereco4bcp+45n100pcBW}
%\includegraphics[width=0.45\textwidth]{ereco4bcp+45n40pcBW}
%PRDv12
\includegraphics[width=0.45\textwidth]{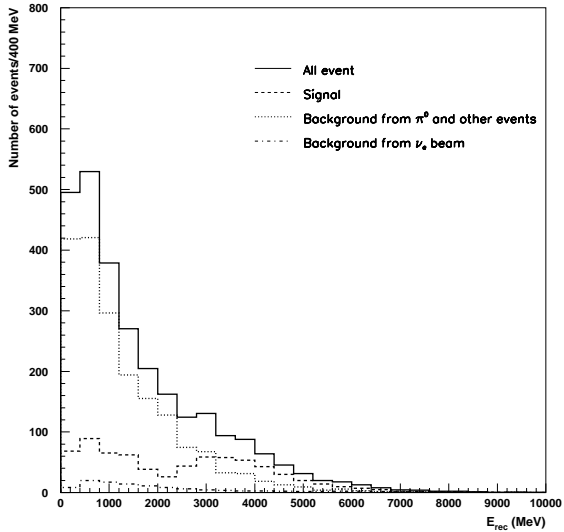}
\includegraphics[width=0.45\textwidth]{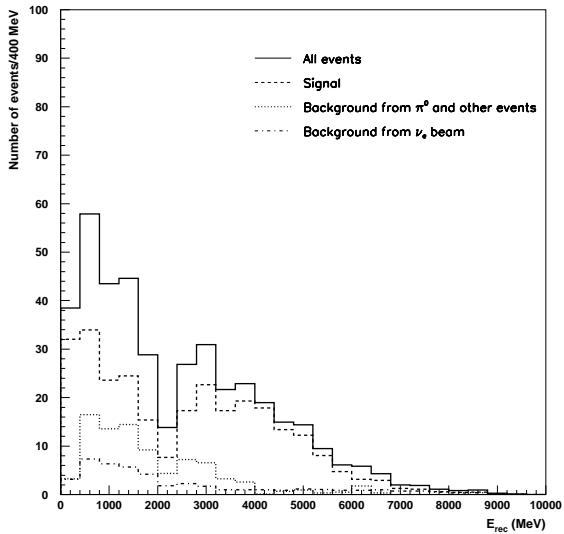}
\caption{\label{fig:ErecDist} Top: The distributions of the
reconstructed neutrino energy with the standard Super-Kamiokande cuts. The
distributions in dashed line are for the signal and those in dotted
(dash-dotted) line are for the background-1(-2). $\delta_{CP}=+45^{\circ}$ and the
baseline is 2,540 km. Bottom: The same distributions except that,
in addition to the standard Super-Kamiokande cuts, the cut on $\Delta logL$ is
applied.}
\end{figure}

In presenting the results, first we illustrate the performance of the
traditional analysis represented by the standard SK-I codes for
comparison.  Figure~\ref{fig:ErecDist} (top) shows the $E_{rec}$
distributions of the signal (dashed line), of the background
(background-1) mostly from NC interactions (dotted line), and of the
irreducible background (background-2) from the $\nu_e$ contamination
in the neutrino beam (dash-dotted line).  The signal is overwhelmed by
the background, especially in the low energy region. In this case, we
find 700 signal events, 1,877 background events from background-1, and
127 background events from background-2.

%\begin{figure}
%\includegraphics[width=0.45\textwidth]{ereco4bcp+45n100pcBW}
%\caption{\label{fig:P100Erec}The distribution of the reconstructed neutrino
%energy with the standard Super-Kamiokande cuts before the cut on $\Delta logL$
%is applied. The distribution in dashed line is for the signal and those
%in dotted (dahs-dotted) line is for the background 1(2). 
%$\delta_{CP}=+45^{o}$ and the baseline is 2,540 km.}
%\end{figure}

On the contrary, if we retain 40\% of the signal events after the cut
on $\Delta \log(L)$ introduced in this paper, the background-1 events
are strongly suppressed as shown in Figure~\ref{fig:ErecDist} (bottom).
We find 280 signal, 87 background-1, and 45 background-2
events.

The analysis presented here uses nine variables to define the
likelihood.  The usefulness of all nine variables is evaluated by
examining how much the signal-to-background (S/B) ratio changes when
each variable is removed in turn from the overall likelihood function.
Table~\ref{tbl:NoEvents} summarizes the numbers of events and the S/B
ratio for each case where only the contribution to background-1 is
considered.
The S/B ratio is always appreciably smaller when one of the nine
variables is removed from the likelihood function than when all nine
are included.  Each variable helps to enhance the signal.

%\begin{figure}
%\includegraphics[width=0.45\textwidth]{ereco4bcp+45n40pcBW}
%\caption{\label{fig:P40Erec+45}The distribution of the reconstructed neutrino
%energy, in addition to the standard Super-Kamiokande cuts, the cut on
%$\Delta logL$ is applied in such a way to retain 40\% of the signal events
%survive by this cut. The distribution in dashed line is for the
%signal and that in dotted (dash-dotted) line is for the background 1(2).
%$\delta_{CP}=+45^{o}$ and the baseline is 2,540 km.}
%\end{figure}

\begin{table}
\caption{\label{tbl:NoEvents}Summary of the numbers of the signal and
background events, and the signal-to-background ratio using events from
background-1 with the cut on $\Delta \log(L)$ to retain 40\% of the signal
events.  The signal-to-background ratio with one variable removed is compared
to the ratio with all nine variables. Note that the numbers of the signal
(background-2) events are not always 280 (45) due to the finite bin size used 
for the distribution of the variable removed.}
\begin{ruledtabular}
\begin{tabular}{c|c|r|c|c}
Variable removed & Sig & Bkg-1 & Bkg-2 & Sig/Bkg-1\\\hline\hline
None & 280 & 87 & 45 & 3.22\\\hline
$\Delta$ log $\pi^{0}$-lh & 281 & 102 & 45 & 2.75\\
Q/E & 281 & 94 & 45 & 2.98\\
log $\pi^0$-lh & 278 & 94 & 47 & 2.98\\
log pid-lh & 277 & 94 & 42 & 2.96\\
E$_{frac}$ & 281 & 98 & 45 & 2.85\\
$m_{\gamma\gamma}$ & 280 & 105 & 45 & 2.66\\
cos$\theta$ & 279 & 101 & 45 & 2.76\\
Cangle & 280 & 98 & 45 & 2.86\\
$\Delta$ log ring-lh & 277 & 95 & 45 & 2.93\\
\end{tabular}
\end{ruledtabular}
\end{table}

\section{$E_{rec}$ distributions and CP-violating phase $\delta_{CP}$}

In the previous section we showed that with a set of appropriate cuts,
the background contribution can be suppressed down to a reasonable level, while
retaining enough statistics for the signal. In this section, we see whether
this set of the cuts is still useful for other values of the CP-violating
phase $\delta_{CP}$. Table~\ref{tbl:NoEventsCP}
lists the numbers of events from the signal, background-1 and background-2 for
various values of $\delta_{CP}$.

% \fixme{Why do the background numbers change here?}
% answer from chiaki: cp changes signal rate, but force 40% eff.

\begin{table}
  \caption{\label{tbl:NoEventsCP}Summary of the numbers of the signal and
    background events for different values of $\delta_{CP}$ using events from
    background-1 (Bkg-1) and background-2 (Bkg-2) with the cut on $\Delta \log(L)$ chosen to retain 40\% of the signal events. The errors are statistical due to
the limited size of the Monte Carlo event sample.}
\begin{ruledtabular}
\begin{tabular}{c|c|r|c}
$\delta_{CP}$ & Sig & Bkg-1 & Bkg-2 \\\hline\hline
% PRDv11
%+135$^\circ$ & 386 & 89 & 50\\
%+45$^\circ$  & 280 & 87 & 49\\
%0$^\circ$    & 197 & 90 & 48\\
%-45$^\circ$  & 159 & 87 & 48\\
%-135$^\circ$ & 263 & 87 & 49\\
% PRDv12
+135$^\circ$ & 386$\pm$6 & 89$\pm$8 & 45$\pm$1\\
+45$^\circ$  & 280$\pm$5 & 87$\pm$8 & 44$\pm$1\\
0$^\circ$    & 197$\pm$4 & 90$\pm$8 & 44$\pm$1\\
-45$^\circ$  & 159$\pm$3 & 87$\pm$8 & 44$\pm$1\\
-135$^\circ$ & 263$\pm$3 & 87$\pm$8 & 45$\pm$1\\
\end{tabular}
\end{ruledtabular}
\end{table}

\section{$E_{rec}$ distributions and baseline}
\begin{figure}
% PRDv11
%\includegraphics[width=0.45\textwidth]{ereco4bcp+45n40pc1480BW}
\includegraphics[width=0.45\textwidth]{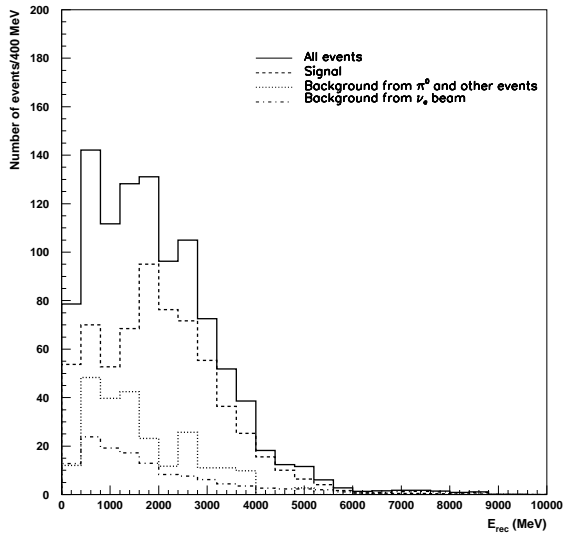}
\caption{\label{fig:P40Erec+45FH}The distributions of the
  reconstructed neutrino energy is shown for events that pass the
  standard SK-I cuts and the cut on $\Delta \log(L)$ such that 40\% of
  the signal events are retained.  The distributions in dashed line
  are for the signal and those in dotted (dash-dotted) line are for
  the background-1 (background-2). $\delta_{CP}=+45^{\circ}$ and the
  baseline is 1,480 km.}
\end{figure}

It is interesting to see how the baseline will change the results of
similar analyses presented in the preceding section. For this study,
the cut on $\Delta \log(L)$ is used again to retain 40\% of signal and
the baseline is assumed to be 1,480 km (Fermilab to Henderson Mine).
Figure~\ref{fig:P40Erec+45FH} shows the $E_{rec}$ distribution of the
signal (dashed line), background-1 (dotted line) and background-2
(dash-dotted) for $\delta = +45^{\circ}$. Table~\ref{tbl:NoEventsCPFH}
lists the numbers of events from the signal, background-1 and
background-2 for different values of $\delta_{CP}$.

\begin{table}
  \caption{\label{tbl:NoEventsCPFH}Summary of the numbers of signal and
    background events for different values for $\delta_{CP}$ using events from
    background-1 (Bkg-1) and background-2 (Bkg-2) with the cut on 
    $\Delta \log(L)$ to retain 40\% of the signal
    events and using a baseline of 1,480 km. The errors are statistical due to
    the limited size of the Monte Carlo event sample.}
\begin{ruledtabular}
\begin{tabular}{c|c|r|c}
$\delta_{CP}$ & Sig & Bkg-1 & Bkg-2 \\\hline\hline
% PRDv11
%+135$^\circ$ & 646 & 238 & 142\\
%+45$^\circ$  & 699 & 233 & 141\\
%0$^\circ$    & 498 & 230 & 140\\
%-45$^\circ$  & 357 & 247 & 142\\
%-135$^\circ$ & 609 & 237 & 142\\
% PRDv12
+135$^\circ$ & 646$\pm$9 & 238$\pm$23 & 133$\pm$2\\
+45$^\circ$  & 698$\pm$11 & 235$\pm$23 & 133$\pm$2\\
0$^\circ$    & 498$\pm$8 & 230$\pm$23 & 132$\pm$2\\
-45$^\circ$  & 356$\pm$6 & 250$\pm$24 & 134$\pm$2\\
-135$^\circ$ & 609$\pm$9 & 237$\pm$23 & 134$\pm$2\\
\end{tabular}
\end{ruledtabular}
\end{table}

\section{Sources of background events}

It is important to know where the background events come from. Information
such as the true energy of neutrinos and nature of interaction of neutrinos
that produce the
background events is very useful for design of the neutrino beam for a VLBNO experiment.

\begin{figure}
% PRDv11
%\includegraphics[width=0.45\textwidth]{enu4bcp+45n40pcBW}
% PRDv12
\includegraphics[width=0.45\textwidth]{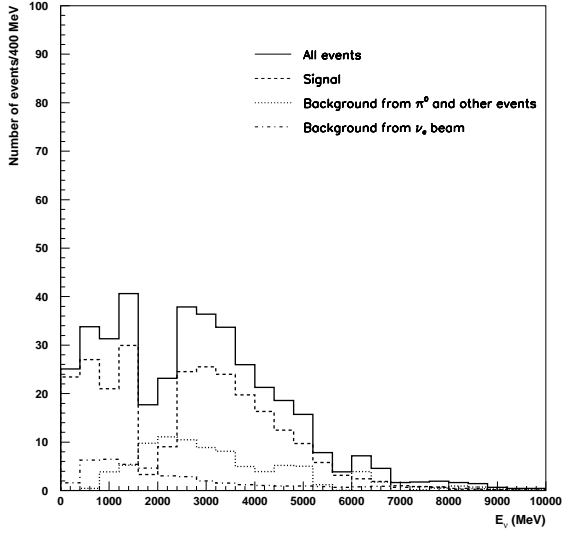}
\caption{\label{fig:enuspec}The distribution of the energies of neutrinos
that produce the signal, background-1 and background-2 events.
In addition to the standard Super-Kamiokande cuts,
the cut on $\Delta \log(L)$ is applied in such a way to retain 40\% of the
signal events that survive by this cut. The distribution in dashed line is 
for the signal and that in dotted (dash-dotted) is for the background-1(-2).
$\delta_{CP}=+45^{\circ}$ and the baseline is 2,540 km.}
\end{figure}

\begin{figure}
% PRDv11
%\includegraphics[width=0.45\textwidth]{enu4bcp+45n40pc1480BW}
% PRDv12
\includegraphics[width=0.45\textwidth]{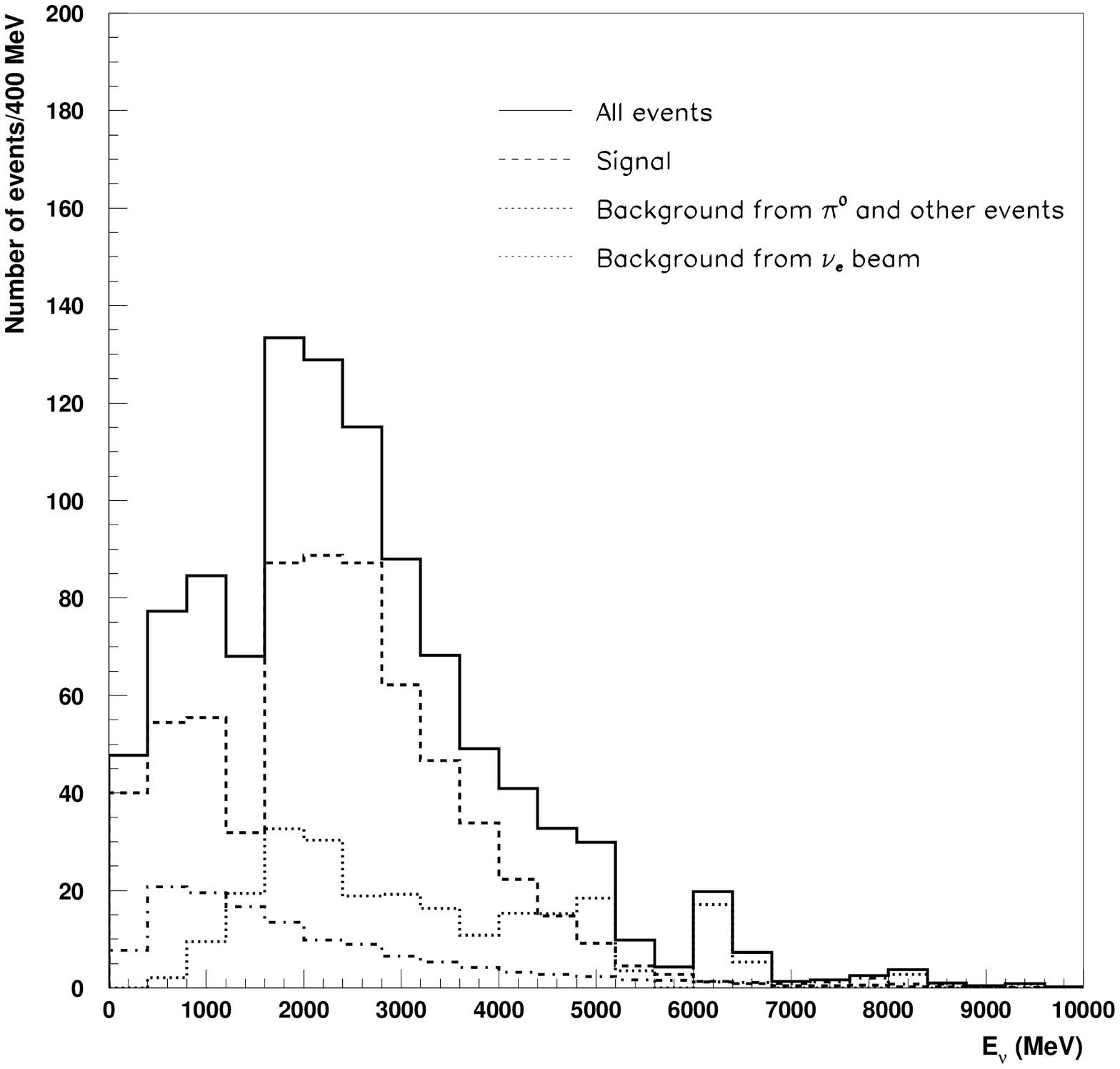}
\caption{\label{fig:enuspec1480}The distribution of the energies of neutrinos
that produced the signal, background-1 and background-2 events.
In addition to the standard Super-Kamiokande cuts,
the cut on $\Delta \log(L)$ is applied in such a way to retain 40\% of the
signal events that survive by this cut. The distribution in dashed line is for 
the signal and that in dotted (dash-dotted) is for the background-1(-2).
$\delta_{CP}=+45^{\circ}$ and the baseline is 1,480 km.}
\end{figure}

Figures~\ref{fig:enuspec}-\ref{fig:enuspec1480} show the neutrino energy
distributions of the signal, background-1 and background-2 events for
the baseline of 2,540 km and 1,480 km, respectively. These events are chosen
with the $\Delta \log(L)$ cut that retains 40\% of the signal events after
the first set of the cuts (the standard SK-I cuts).

As mentioned earlier, for the most of the results presented in this
report, we apply the cut on the neutrino energy at 10 \gev{}. To justify this
cut, we checked to see how much more the background contribution would increase
if we allowed events produced by neutrinos whose energies were greater than
10 \gev{} and less than 15 \gev{}. 
For these additional events we fixed the weight value to that used at
10 \gev{}.
The atmospheric $\nu_e$ flux times the $\nu_e$ cross section decreased
by about 60\% while that for the wideband $\nu_{\mu}$ beam by about
50\% in this energy region. Therefore the atmospheric $\nu_e$ and
$\nu_{\mu}$ spectrum are only slightly softer than the wideband
$\nu_{\mu}$ spectrum from 10 \gev{} to 15 \gev{}. This argument,
therefore, gives a reasonable estimate of the extra contribution from
high energy neutrinos. All the neutrino oscillation parameters are the
same as in the case with $\delta_{CP}=+45^\circ$, the baseline of
2,450 km, and the 40\% $\Delta \log(L)$ cut. The numbers of the
signal, background-1, and background-2 events accepted are 281, 91 and
45, respectively, which should be compared with 280 for the signal, 87
for the background-1 and 45 for the background-2 with the neutrino
energy cut at 10 \gev{}. Thus the percent increases in the number of
the signal, background-1 and background-2 events are 1\%, 4\%, and
0\%, respectively.
 
Tables~\ref{tbl:intmode} and \ref{tbl:intmode1480} summarize all
neutrino interactions that produce candidate events passing all
cuts with 40\% efficiency for the baseline of 2,540 km and 1,480 km,
respectively.  In these tables, $\pi^{0}$,$\pi^{\pm}$, and $n\pi$
stand for single $\pi^0$, single $\pi^{\pm}$ and multiple $\pi$
production, respectively.  The co$\pi$ label stands for $\pi$
%% clear up by CY on 5/3/10
%production via coherent interaction with the Oxygen nucleus and
%``others'' includes eta and kaon production as well as deep inelastic
%scattering (DIS). Note that in the tables there is no contribution
production via coherent interaction with the oxygen nucleus. The
``others'' includes eta and kaon production as well as deep inelastic
scattering (DIS). However, for the background, the label ``$n\pi$'' includes
DIS. Note that in the tables there is no contribution
From NC interaction for signal events by definition.  CC QE
contributions to background-1 are due to $\nu_\mu$ interactions.
% Addition on 1/8/11 for PRDv12
Since the cross section of the coherent pion production is known only to
an accuracy of $\pm$30\% (see reference \cite{ref:ATMFull}), we estimate the 
effect of this error on our background-1 contribution by varying the cross 
section by $\pm$30\%. For $\delta_{CP}= +45^\circ$ with the baseline of 2,450 km
$\pm$30\% changes in this cross section result in changes in the numbers of 
background-1 events by $\pm$3\%.

\begin{table}
\caption{\label{tbl:intmode}The percent contributions from events
produced by different interactions for signal events and for the background-1
events are summarized for $\delta_{CP}=+45^\circ$ with the cut on
$\Delta \log(L)$ to retain 40\% of the signal events and with the baseline of
2,540 km.}
\begin{ruledtabular}
\begin{tabular}{l|r|r|r|r|r|r}
\multicolumn{1}{c|}{Interaction} &
\multicolumn{6}{c}{$E_{rec}$ range (\gev{})} \\\hline\hline
Sig  & 0.0-0.5 & 0.5-1.0 & 1.0-1.5 & 1.5-2.0 & 2.0-3.0 & 3.0- \\\hline
CC QE  & 86\% & 79\% & 63\% & 82\% & 28\% & 46\%\\
CC $\pi^0$     &  2\% &  3\% &  4\% &  2\% &  5\% &  6\%\\
%% This line is the result of consolidation of two lines below
CC $\pi^{\pm}$ &  11\% & 15\% & 28\% &  13\% & 37\% & 30\%\\
%CC $\pi^{\pm}$ &  11\% & 15\% & 27\% &  12\% & 37\% & 30\%\\
%CC co$\pi^+$   & 0\%  & 0\% & 1\% & 0\% & 2\% & 2\% \\
CC $n\pi$      &  1\% &  3\% &  3\% &  4\% & 25\% & 14\%\\
CC others      &  0\% &  0\% &  1\% &  0\% &  3\% &  2\%\\\hline
\hline
Bkg-1& 0.0-0.5 & 0.5-1.0 & 1.0-1.5 & 1.5-2.0 & 2.0-3.0 & 3.0- \\\hline
CC QE          &  7\% &  5\% &  6\% &  0\% &  0\% &  0\%\\
CC $\pi^0$     &  0\% &  1\% &  4\% &  6\% &  4\% &  0\%\\
CC $\pi^{\pm}$ &  2\% &  5\% &  0\% &  1\% &  0\% &  0\%\\
CC $n\pi$      &  0\% &  0\% &  3\% &  6\% & 0\% &  0\%\\
CC others      &  0\% &  0\% &  0\% &  0\% &  21\% &  3\%\\
NC $\pi^0$     & 23\% & 53\% & 60\% & 59\% & 18\% &  0\%\\
NC $\pi^{\pm}$ & 64\% & 10\% &  6\% &  0\% &  0\% &  0\%\\
%% This line is the result of consolidation of old NC npi and old NC DIS
NC $n\pi$      &  0\% & 13\% &  5\% & 16\% & 55\% & 92\%\\
%NC $n\pi$      &  0\% & 13\% &  0\% & 9\% & 0\% & 0\%\\
NC co$\pi^0$ & 5\% & 14\% & 16\% & 12\% & 2\% & 5\%\\
%NC DIS           & 0\% & 0\% & 5\% & 7\% & 55\% & 92\% \\
NC elastic           & 0\% & 0\% & 1\% & 0\% & 0\% & 0\% \\
\end{tabular}
\end{ruledtabular}
\end{table}

\begin{table}
\caption{\label{tbl:intmode1480}The percent contributions from events
produced by different interactions for the signal events and for the 
background-1 events are summarized for $\delta_{CP}=+45^\circ$ with the cut on
$\Delta \log(L)$ to retain 40\% of the signal events and with the baseline of
1,480 km.}
\begin{ruledtabular}
\begin{tabular}{l|r|r|r|r|r|r}
\multicolumn{1}{c|}{Interaction} &
\multicolumn{6}{c}{$E_{rec}$ range (\gev{})} \\\hline\hline
Sig  & 0.0-0.5 & 0.5-1.0 & 1.0-1.5 & 1.5-2.0 & 2.0-3.0 & 3.0- \\\hline
CC QE          & 82\% & 77\% & 52\% & 42\% & 54\% & 53\%\\
CC $\pi^0$     &  3\% &  3\% &  6\% &  7\% &  4\% &  5\%\\
%% This line is the result of consolidation of two lines below
CC $\pi^{\pm}$ & 14\% & 18\% & 30\% & 36\% & 30\% & 30\%\\
%CC $\pi^{\pm}$ & 14\% & 17\% & 30\% & 35\% & 28\% & 28\%\\
%CC co$\pi^+$ & 0\% & 0\% & 0\% & 1\% & 2\% & 2\% \\
C $n\pi$      &  1\% &  2\% & 12\% & 13\% &  11\% & 10\%\\
CC others      &  0\% &  0\% &  0\% &  2\% &  1\% &  2\%\\\hline
\hline
Bkg-1& 0.0-0.5 & 0.5-1.0 & 1.0-1.5 & 1.5-2.0 & 2.0-3.0 & 3.0- \\\hline
CC QE          &  8\% &  4\% &  1\% &  3\% &  0\% &  0\%\\
%CC $\pi^0$     &  0\% &  2\% &  6\% &  4\% &  1\% &  0\%\\
%CC $\pi^{\pm}$ &  6\% &  6\% &  2\% &  1\% &  0\% &  0\%\\
% Reverse the numbers for 2.0-3.0 7/22/10
CC $\pi^0$     &  0\% &  2\% &  6\% &  4\% &  0\% &  0\%\\
CC $\pi^{\pm}$ &  6\% &  6\% &  2\% &  1\% &  1\% &  0\%\\
CC $n\pi$      &  0\% &  0\% &  1\% &  1\% &  0\% &  0\%\\
CC others      &  0\% &  0\% &  0\% &  4\% &  2\% &  7\%\\
NC $\pi^0$     & 23\% & 58\% & 61\% & 29\% & 23\% &  0\%\\
NC $\pi^{\pm}$ & 59\% &  5\% &  6\% &  0\% &  0\% &  0\%\\
%%% This is the result of consolidation of NC npi and NC DIS
NC $n\pi$      &  0\% & 11\% &  5\% & 23\% & 65\% & 93\%\\
%NC $n\pi$      &  0\% & 11\% &  0\% & 23\% & 0\% & 0\%\\
NC co$\pi^0$  & 4\% & 14\% & 17\% & 35\% & 9\% & 0\% \\
%NC DIS        & 0\% & 0\% & 5\% & 0\% & 65\% & 93\% \\
NC elastic         & 0\% & 0\% & 1\% & 0\% & 0\% & 0\% \\
\end{tabular}
\end{ruledtabular}
\end{table}

\section{Detector size and granularity}

It is interesting to see what the effect of the detector size has on the
performance of \polfit{}. Although the results reported so far
are based on the analyses of the Monte Carlo events generated for the
Super-Kamiokande detector with 40\% PMT coverage, we can make some assessment
of the effect of a larger, more granular detector by imposing a cut on the distance to the PMT surface from
the $\pi^0$ production point in the direction of the $\pi^0$ (\dmin). For
this study we use single $\pi^0$ events produced by NC
interactions.  
When the
$\pi^0$ energy is about 1 \gev{}, the minimum opening angle of
two photons is about 20$^\circ$ and less at higher energies. 
As \dmin{} gets larger, the number of PMTs that detect Cherenkov
photons increases (improved granularity) which can help to resolve light
patterns.  Two
negative impacts compete with this effect.  The Cherenkov cone is more
spread out so less light is seen by any one PMT.  The light must
travel further and is thus more subject to absorption and scattering
(attenuation) in the water.  These both lead to a decrease in the
number of detected photons per PMT which may degrade the pattern of light
that the Cherenkov cone produces.  In the first case, since
information is not lost and is just spread out to more PMTs the
problem can likely be handled in improvements to reconstruction codes.
In the second case, some information about the original Cherenkov
light emitting particles is unrecoverable.

\begin{figure}
\includegraphics[width=0.45\textwidth]{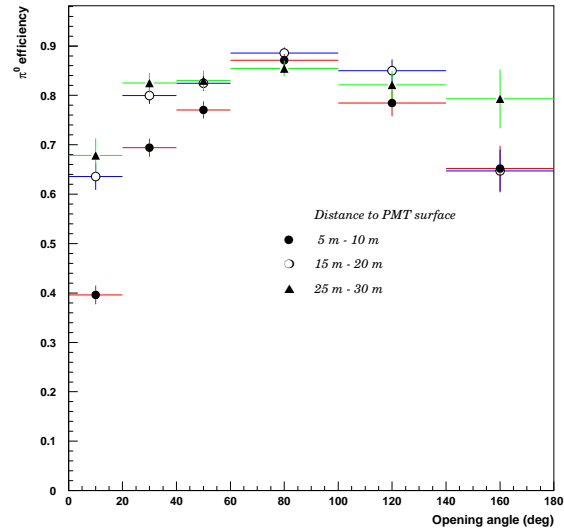}
\caption{\label{fig:pi0size}The $\pi^0$ detection efficiency as a function
of the two photon opening angle for three ranges of the distance from the
$\pi^0$ production vertex to the closest PMT surface in the direction of
$\pi^0$.}
\end{figure}

Figure~\ref{fig:pi0size} shows the $\pi^0$ detection efficiency as a function
of the opening angle for \dmin{} ranges: 5 m - 10 m, 15 m - 20 m, and 25 m - 30 m and includes \polfit{} information.
It is clearly seen that when the opening angle is smaller (less than 60$^\circ$),
the efficiency is improved as the distance to the PMT surface in the $\pi^0$
direction increases. Note that above the opening angle of 60$^\circ$, the $\pi^0$
detection efficiency seems more or less independent of \dmin{}. 
Because of the finite size of the detector and at such large opening
angles selecting events based on \dmin{} becomes a less capable means
of emulating a larger detector.

Nevertheless, this indicates that the granularity of the detector in
terms of the PMT density is an important factor to improve the $\pi^0$
detection efficiency.
It also appears that the effect of light attenuation is not a major
issue at least for Cherenkov light traveling up to 30-40 m
in SK-I water which has an attenuation length of 80-100 m (depending
on wavelength).

Therefore for the same PMT coverage using the same PMTs, a larger
detector than SK-I will perform better, on average, to reconstruct
$\pi^0$.  Limitations may be expected once the typical path length for
Cherenkov light approaches attenuation length.

A similar study can be done by looking at how the S/B ratio varies as
a function of \dmin{} for other event types.  
%For example,
%Table~\ref{tbl:intmode1480} shows there is a significant contribution
%from $\nu_e$ CC-$\pi^\pm$ events ($\nu_{e}+N\rightarrow e^{-}+N'+\pi^{\pm}$)
%to the sample classified as signal.
In this case we look at the S/B ratio for values of \dmin{} for the
primary e-like ring.  For \dmin{} $>20$ m the S/B ratio changes from
the average of 1.4 to 3.8 for events with $E_{rec}\leq$ 1.2~\gev{}, and
for events with 2~\gev{} $\leq E_{rec} <$ 4 \gev{} the S/B ratio essentially
stays the same.

The large improvement in the S/B ratio for events
with $E_{rec}\leq$  1.2 \gev{} results from an increase in the number of PMTs
(pixels) in a Cherenkov ring. This improvement is significant as in the energy
region $E_{rec}\leq$ 1.2~\gev{} the contribution from the NC events
is reduced to a level as low as that from the irreducible background. This
improvement, however, is not realized for
events with 2~\gev{} $\leq E_{rec} <$ 4~\gev{} presumably because in this energy
region multi-pion events are the major background. 

This observation is also
true if the minimum distance \dmin{} is set at 10 or 15 m, although the
improvement in the S/B ratio is less than the case of the 20-m cut.
For a SK-I sized detector, this 20-m cut reduces the number
of the signal events by 41\%. However, if the detector is larger, this loss
of efficiency can be greatly reduced. In other words, for a given detector
size, the finer granularity, not necessarily the number of Cherenkov
photons collected by individual PMT, improves the S/B ratio.

\section{Conclusion}
% PRDv12
For the baseline of 2,540 (1,480) km the detection efficiency of the signal
events using the SK-I cuts only is found to be 0.361$\pm$0.003 (0.373$\pm$
0.003) and the final efficiency with the further 40\% cut on the 
log-likelihood ratio is found to be 0.145$\pm$0.002 (0.149$\pm$0.002). The 
detection efficiency of the background-1 events using the SK-I cuts only is
found to be 0.054$\pm$0.001 (0.059$\pm$0.001) and the final efficiency with the
further 40\% cut on the log-likelihood ratio is found to be 0.0025$\pm$0.0002
(0.0026$\pm$0.0003). The 40\% cut on the log-likelihood should be considered
as a guidance and the actual cut should be optimized after a detailed study of
figure of merit that depends on the experimental design including the
neutrino beam properties of a given experiment.

We have demonstrated that a large water Cherenkov detector can be used to
detect efficiently the signal events by $\nu_e$s from the neutrino oscillation
$\nu_\mu\rightarrow\nu_e$ while keeping the background at a reasonably low
level in a VLBNO experiment for the baseline
of over 1,400 km with a wideband beam.

\begin{acknowledgments}

  % We wish to thank the Super-Kamiokande collaboration, especially the
  % atmospheric neutrino and proton decay group, who have been making
  % never-ending efforts to write the world's best Monte Carlo event
  % generator and analysis software for water Cherenkov detectors.
  % \fixme{This is two flowery.  Can you suggest an alternative?}
  % \fixme{We also need to acknowledge whatever DOE grant number(s) this
  %   work was done under.}

% Added 7/22/10
  The authors gratefully acknowledge the work of the Super-Kamiokande
  collaboration in developing the most of tools used for this analysis.
  However, the result and its interpretation are responsibility of the
  authors of this paper. The work is partially
%  We thank the Super-Kamiokande collaboration to allow us to use their
%  Monte Carlo event samples and analysis codes to conduct the work
%  summarized in this report. In particular, a special thanks goes to
%  the atmospheric neutrino and proton decay group who has developed
%  most of analysis codes used in this study. The work is partially
  supported by funding from Stony Brook University Office of the Vice
  President for Research, DOE grant DEFG0292ER40697 at Stony Brook 
  University and DOE contract DE-AC02-98CH10886 at Brookhaven 
  National Laboratory, 
  and the City University of New York PSC-CUNY Research
  Award Program at Borough of Manhattan Community College/the City
  University of New York.

\end{acknowledgments}

\end{document}